\documentclass{mpe_report}

\usepackage{psfig,graphicx,epsfig}
\usepackage{color}
\usepackage{amsmath,amssymb,epic,eepic,array}

\begin{document}

\pagenumbering{arabic}
\setcounter{page}{197}

 \renewcommand{\FirstPageOfPaper }{197}\renewcommand{\LastPageOfPaper }{200}
\title{Pulsar Physics without Magnetars}
\author{W.Kundt}  
\institute{Argelander Institute of Astronomy, Bonn University, Auf dem H\"ugel 71, D-53121 Bonn, Germany}
\maketitle{}

\begin{abstract}
Almost 40 years after the discovery of pulsars -- and despite a plethora of secured data on them 
-- pulsar theory is still beset by a number of fundamental inconsistencies. In this short 
contribution, I will argue that (i) magnetars do not exist, (ii) (ordinary) pulsars turnoff (or `die')
when their wind pressure falls short of keeping the CSM at a safe distance, exceeding $10^{15}$cm,
whereupon they can mimic magnetars, (iii) msec pulsars are born fast (in core-collapse SNe), and are 
much older inside globular clusters than outside of them, (iv) neutron-star corotating magnetospheres 
can oscillate almost in resonance with their spin frequency, giving rise to pulse drifting, and to QPOs
of accreting binary X-ray sources, and (v) the dying pulsars are the dominant sources of the cosmic
rays, and of the GRBs.       
\end{abstract}

\section{Four constraints on pulsar physics}

Here are four theses on pulsars which I favour for more than 13 years over alternative ones, for reasons given 
subsequently, and which I will use as assumptions in the rest of this communication. They will lead to new 
insights -- hopefully correct ones -- into the many important roles which pulsars play in the Galaxy. They are:

$\bullet$ Pulsars blow strong, leptonic, extremely relativistic winds.

$\bullet$ Pulsars die statistically at a (spindown) age of $10^{6.4}$yr.

$\bullet$ Pulsar magnetic fields are dipoles stabilized by toroidal bandages.

$\bullet$ Pulsar surfaces are covered by (soft X-ray) hot pair coronae.

The first of these four theses is gleaned from the fact that at least 17 (nearby) pulsars have been seen to blow 
bowshocks into their CSM, of radii between $10^{15}$cm and $10^{18}$cm, mapped at H$\alpha$, X-rays, radio, and/or 
even broadband (Kundt, 1998). Ram-pressure-balance estimates imply wind densities some $\xi=10^4$ times the (shunting) 
Goldreich-Julian density escaping at relativistic speeds, i.e. very strong winds when launched by the unipolar-induction 
electric voltage (Kundt \& Schaaf 1993). I consider such strong winds incompatible with polar-cap sparking, or with
outer gaps in the magnetosphere. Note that bowshocks can be missing, cf. Hui \& Becker (2006): probably in underdense 
regions of the Milky Way, if the latter floats on pair plasma as the volume-filling medium (Kundt 2004, p.35).

A statistical pulsar age of $10^{6.4}$yr can be read off Fig. 1: the proportionality N $\sim \tau$ drops exponentially 
beyond this spindown age. In my 2005 contribution to the Berlin-Adlershof meeting, I have revived an ancient explanation:
The pulsar wind cavity blown into the CSM -- against the neutron-star's gravitational attraction -- can be shown to 
exceed $10^{14.9}$cm$/T_3$ in radius for an effective ambient temperature in units of $10^3$K, ($T_3 := T/10^3 K$). 
As will be reviewed in the next section, this critical minimum size can no longer be sustained when the pulse period 
grows beyond several sec, Eq 1 (which depends on the surroundings), whereupon the CSM avalanches down onto the pulsar's 
magnetosphere and throttles it, in the form of a very-low-mass accretion disk which cuts deeply into it, down to the 
corotation radius. In my understanding, such `throttled pulsars' spin down fast, due to their increased magnetic torque 
which scales as $r^{-3}$ with increasing confinement $r^{-1}$, and can be confused with a magnetar. It can appear as an 
AXP, or SGR, or as a `stammerer' (= rotating radio transient = RRAT); its luminosity is mainly powered by accretion.

\begin{figure}
\centerline{\psfig{file=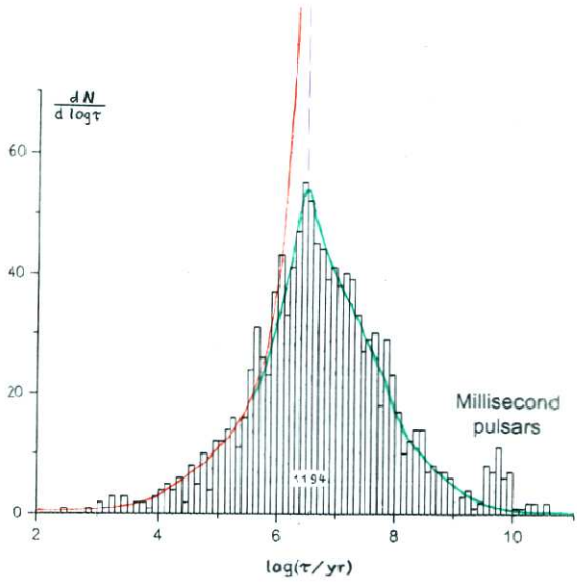,width=8.0cm} }
\caption{N = 1194 pulsars plotted linearly w.r.t. their logarithmic spindown age, dN/dlog$\tau$ vs log($\tau$/yr), 
$\tau := P/2\dot P$. For a stationary age distribution, the upper envelope would rise exponentially, as drawn in - 
both solid and broken - for two extreme interpretations of the noise. Clearly, there is an increasing deficit of 
detected pulsars for $\tau \ge 10^{6.4}$yr. The small bump of ms pulsars, of spindown ages between $10^{9.5}$yr and 
$10^{10}$yr, may be due to those in globular clusters.
\label{image}}
\end{figure}

Pulsar magnetic fields tend to be approximated by dipoles even though we know since Flowers \& Ruderman (1977) that 
a dipole field inside a fluid star is dynamically unstable. A pulsar, born inside a core-collapse SN, receives a 
stabilizing toroidal magnetic bandage whose presence can be described by an expansion in terms of odd-order multipoles 
(Kundt 1998, 2004), see Fig 2. The higher multipoles guarantee that magnetic curvature radii near the surface are 
smaller than, or comparable to the stellar radius, that surface magnetic field strengths are (larger than for the dipole), 
large enough for the Erber mechanism (to convert hard photons into e$^{\pm}$-pairs), and that polar caps are 
correspondingly small.     
    
\begin{figure}
\centerline{\psfig{file=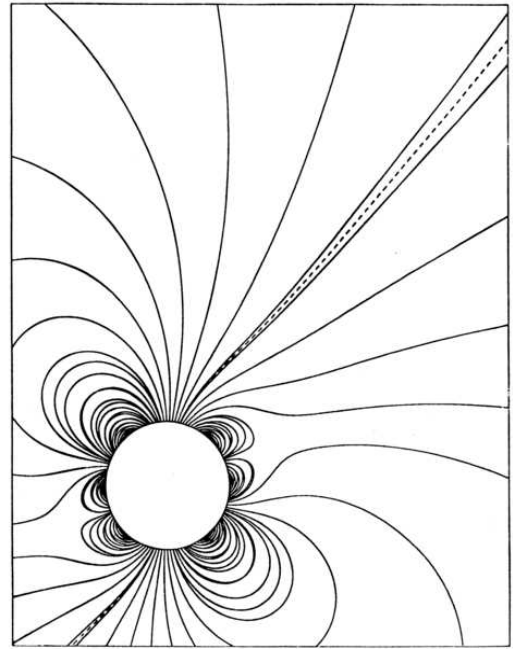,width=7.0cm} }
\caption{Plausible pulsar magnetosphere, obtained by adding 6 times a normalized octupole to a dipole 
inclined by 40 deg, from Chang (1994).
\label{image}}
\end{figure}

Standard theory predicts huge electric (unipolar induction) fields at neutron-star surfaces, strong enough to create 
electron-positron pairs wherever unscreened through distances exceeding one cm, corresponding to strongly sub-nsec 
delay times (of screening). Time-dependent descriptions of the wind-formation problem -- expected in an AC mode -- 
predict strong outgoing relativistic electric currents near the polar caps, plus weaker ingoing electric currents, and 
similar -- perhaps somewhat weaker -- currents across the whole neutron-star surface. The ingoing currents bombard the surface, and create pairs. 
Pair formation will saturate when their creation is compensated by their annihilation, which should happen when this pair 
corona gets thick w.r.t. the annihilation cross section (in the strong magnetic field). The temperature of the corona 
was obtained in Kundt \& Schaaf (1993) through balancing the bombardment power by outgoing blackbody radiation, of order 
$10^{6.5}$K, which implies a corona scale height of order $10^2$m. Such a corona eliminates the work-function problem for 
wind formation -- because both types of charges are freely available -- and implies blackbody spectra from the surface 
whenever the corona is optically thick to Thomson scattering, i.e. whenever the (magnetic) pair-annihilation cross 
section is large compared to the Thomson cross section. In (Kundt 2002), I estimated their ratio as $\ga$ 10. Such 
blackbody spectra have meanwhile been secured for the seven musketeers (Frank Haberl, Roberto Turolla, these proceedings).

\section{The throttled pulsars can replace the magnetars}
  
Pulsar wind formation requires a cavity around the magnetized rotator, blown by its outgoing relativistic flux (into its 
circumstellar medium, CSM), cf. Kundt (2004). The wind pressure $p = L/4 \pi r^2 c$ at distance $r$ scales as the power $L$ which is 
thought to almost equal the pulsar's spindown power $L \la - I \Omega d \Omega / dt$. The CSM feels the neutron star's 
gravity. Its weight at radial distance $r$ overcomes the wind pressure $p \sim \Omega ^4/r^2$ when the angular velocity 
$\Omega =: 2 \pi /P$ drops below a critical value given by

\begin{equation}
P \ge 8s(\mu_{31} T_3 / \sqrt{p_{-12.3}})^{1/2} ,
\label{eq:1}
\end{equation}

\noindent
where $\mu$ is the star's magnetic dipole moment, $\mu _{31} := \mu/10^{31}$G cm$^{3}$, and $T$ is a typical value for 
the temperature of the CSM, thought to be multi-component, with (cold) HI filaments embedded in relativistic pair 
plasma. At the epoch of suffocation, or throttling, the cavity radius $r$ takes its minimum value

\begin{equation}
r_{cav} \ge G M m / 2 k T = 10^{14.9}cm / T_3 .
\label{eq:2}
\end{equation}

\noindent
During the throttling event, the CSM avalanches down towards the pulsar, and quenches its corotating magnetosphere. If 
homogeneous, the encaving plasma confines the dipole such that its magnetic torque $\approx r^3 <B_r B_{\phi}>$ rises 
as $r^{-3}$ with increasing confinement, implying faster spindown than before -- in proportion to $(c/\Omega r)^3 \la 
10^{13.5}$ -- and stronger emission.

More realistically, most of the impacting CSM will form a low-mass accretion disk (of mass $\la 10^{-6} M_{\odot}$) 
around the pulsar -- because of its 
large angular-momentum excess during collapse -- whose inner edge interacts even more strongly with the corotating 
magnetosphere, and generates cosmic rays. Statistically, these low-mass accretion disks will tend to be oriented 
perpendicular to the Milky-Way plane, when the pulsar oscillates `up and down' through the Galactic disk. Figure 3 
sketches the scenario before and after infall, on length scales of $10^{15}$cm and $10^8$cm, respectively.

\begin{figure}
\centerline{\psfig{file=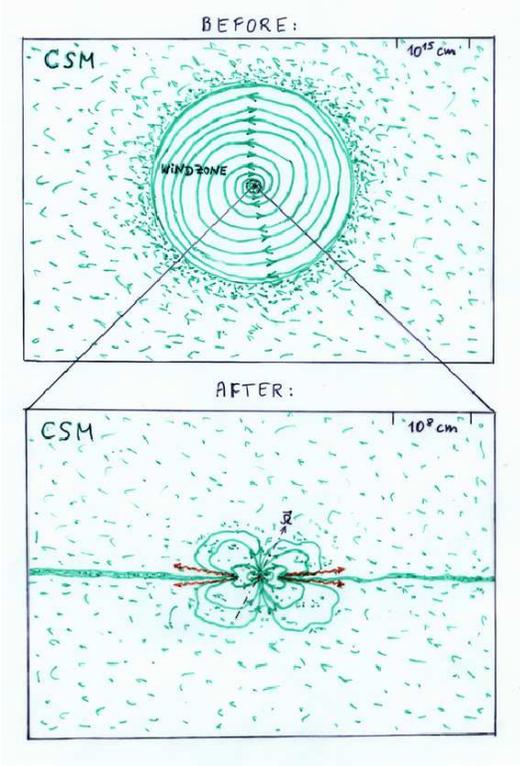,width=7.0cm} }
\caption{Cartoon sketching a pulsar's suffocation, on two scales: once the heavy `atmosphere' of its windzone quenches 
it, by free-falling down under angular-momentum conservation, it forms a low-mass accretion disk cutting deeply into 
its corotating magnetosphere, resembling a relativistic grindstone (at its inner edge). CR and impact emissions will 
be preferentially in the plane of the (inner) disk.
\label{image}}
\end{figure}

Historically, magnetars were invented by Duncan \& Thompson (1992) to explain the (rather isotropically distributed) 
gamma-ray bursts by Galactic halo sources, with a subsequent shift towards certain pulsed isolated neutron-star sources 
in rapid spindown, the SGRs and AXPs, see Thompson \& Duncan (1996). The magnetars, if realistic, would form a 
detached population from all the other compact pulsators, with internal magnetic field strengths of order $10^{17}$G 
which are difficult to anchor (by the neutron-star core fluid). One of them, SGR 1806-20, showed an upward jump in 
$\dot P$ during a glitch late in 2004, which would have corresponded to a sudden spontaneous increase in $\mu$! I 
prefer to think that Duncan \& Thompson's sources are the dying pulsars.

This new class of sources -- throttled pulsars -- offers plausible interpretations for the following characteristic 
properties of them (Mereghetti et al 2002):

(1) They are isolated neutron stars, with spin periods $P$ between 5 s and 12 s, and similar glitch behaviour 
correlating with X-ray bursts.

(2) They are soft X-ray sources, hotter than pulsars of the same spindown age by a factor $\ga$ 3 -- explained 
as due to magnetospheric interactions with the throttling CSM 
and/or mild accretion -- yet mostly with no pulsed coherent radio emission (Camilo et al. 2006).

(3) Their spindown is rapid, $\tau = 10^{4 \pm 1}$yr, despite ongoing accretion.

(4) Their expected number is the number of observed pulsars near the peak of their distribution (w.r.t. age), some 
$10^3$, reduced by the ratio of their respective (shortened) spindown times ($10^2$), i.e. some 10.

(5) They derive their power ($\la 10^{36}$erg/s) from accretion, whose implied spinup is overcompensated by 
magnetospheric braking.

(6) They are -- at the same time -- the dominant sources of the cosmic rays, and of the (extraterrestric) $\gamma$-ray 
bursts (Kundt 2004).

(7) They are often (some 50\%) found near the center of a pulsar nebula (Gotthelf et al 2000).

\section{The problem of the msec pulsars}

The ms pulsars are often called `recycled' -- rather than born fast -- even though no single progenitor has ever been 
identified: all the accreting X-ray binaries are found in quasi-steady equilibrium between spinup and spindown (when 
observed long enough, for more than a decade), and even though their masses (after accretion!) show no increase, even 
for the fastest of them; see also (Kundt 2004). Recycling estimates have been made with neglect of the braking torques.

The biggest problem with the ms pulsars is their overabundance in globular clusters; it shows up as an extra bump in 
Fig 1, despite the small mass of the system of globular clusters compared with the (mass of the) Galaxy, and despite
the low escape velocity from globular clusters, lower than typical birth velocities of (ordinary) pulsars. This 
conundrum can be resolved when ms pulsars inherit lower birth velocities than their slower cousins (because of weaker 
magnetic dipole moments), and when their (true) ages in the Galaxy are some $10^3$ times shorter than their spindown 
ages, whilst comparable to them in the globular clusters because of a quasi-weightless CSM there. No simpler solution 
(of this conundrum) has reached my mind yet.

\section{Drifting subpulses and X-ray QPOs}

As mentioned at this Seminar, the phenomenon of drifting subpulses appears to be quite general; and its usual explanation 
by E x B-drifting sparks around polar caps may be in mild conflict with their strong observed winds. How about 
oscillating magnetospheres? Here is a sketchy estimate of their oscillation frequency: it equals their rotation frequency, 
at least approximately, hence can be easily excited at resonance. The tiny inertia of the outgoing pulsar wind modulates 
this resonance slightly. At the same time, oscillating magnetospheres can explain the so far ill-understood QPO phenomenon 
of the accreting binary X-ray sources, both neutron-star and BHC binaries.

The oscillation equation for a rotator of moment of inertia $I$, torque $T$, torque gradient w.r.t. angle $\phi$ equal to 
$T' := dT/d\phi$ reads   
 
\begin{equation}
I (\delta \phi)^{\dot{}\ \dot{}} = -T' \delta \phi.
\label{eq:3}
\end{equation}

\noindent
Here $(\delta \phi)^{\dot{}\ \dot{}} = -\omega ^2 \delta \phi$ (for an oscillation angular frequency $\omega$), and $I$ = $Mr^2$ 
for an inertial mass $M = E/c^2$ of the electromagnetic field of the magnetosphere, and inertial radius $r = c/\Omega$ at 
the speed-of-light cylinder, so that we get

\begin{equation}
\omega /\Omega = \sqrt{T'/T} \approx 1
\label{eq:4}
\end{equation}

\noindent
because the outgoing power of the corotating magnetosphere can be alternatively expressed as $T \Omega$, or as the surface 
integral $S = E \Omega$ over the outgoing Poynting flux, whence $E = T$, and because $T'/T = dlnT/d\phi \approx 1$. 

This estimate shows that unloaded magnetospheres can oscillate in near-resonance with their rotation, excited, e.g., by 
their variable load, so that all pulsars are expected to (slightly) drift.

\section{The conundra of the CRs, and of the GRBs}

The origins of the cosmic rays (CRs), and of the (quarter-daily) gamma-ray bursts (GRBs) are among the hardest conundra of 
present-day astrophysics: where do these ultrahard `radiations' come from? My own conviction -- now dating back 30 years 
-- has been their generation in Galactic neutron stars, in a transrelativistic slingshot mode. Because here we deal with 
the deepest-known, strongly variable potential wells, so that neither their energetics (up to $10^{20.5}$eV per proton!) 
nor their rapid fluctuations ($\la 10^{-3.7}$s for GRBs), nor their high repetition rates pose problems to theorists 
(Kundt 2004).

The conundra formed with the findings of their isotropic arrival directions -- for the CRs only at the highest-energy end 
of their spectrum ($\ga 10^{19}$eV), where they propagate almost like photons -- yet with the occasional occurrence of 
repeaters, seemingly contradicting a Galactic disk population. But precisely this property is expected for the (large) 
population of dying (throttled) pulsars, whose low-mass accretion disks are oriented roughly at right angles to the 
Galactic plane, and whose radiations are therefore similarly directed out of the Galactic plane. In this way, 
there is a first-order compensation between a latitude-dependent increase of sources on approach of the plane, and an 
equally $\theta$-dependent decreasing probability, $\sim sin(\theta)$, for us to be in the beam, resulting in an (almost) 
isotropy of arrivals.

\section{Conclusions}
Pulsars form a great astrophysical testing ground.
  
\vskip 0.4cm

\begin{acknowledgements}
My warm thanks go to Hans Baumann and Gernot Thuma for the manuscript, and to G\"unter Lay for the electronic data 
handling.
\end{acknowledgements}

              \clearpage

\end{document}